\documentclass[aps,prb,10pt,twocolumn,superscriptaddress]{revtex4-1}

\usepackage{graphicx}
\usepackage{float}
\usepackage{amsmath}
\usepackage{textcomp, gensymb}
\usepackage{upgreek}
\usepackage{xcolor}
\usepackage{lipsum}

\definecolor{paperblue}{HTML}{2B54B0}
\usepackage{hyperref}
\hypersetup{colorlinks=true,urlcolor=paperblue,linkcolor=black,citecolor=paperblue}

\usepackage[final]{microtype}
\newcommand{\EFfilms}{1}
\newcommand{\EFhistograms}{2}
\newcommand{\EFplaneorientation}{3}
\newcommand{\EFfullxrd}{4}
\newcommand{\EFtwinning}{5}
\newcommand{\EFtem}{6}
\newcommand{\EFfits}{7}

\begin{document}

\title{Single-crystalline PbTe film growth through reorientation}

\author{Jason Jung}\thanks{These authors contributed equally to this work.}
\affiliation{Department of Applied Physics, Eindhoven University of Technology, 5600 MB Eindhoven, The Netherlands}
\author{Sander G. Schellingerhout}\thanks{These authors contributed equally to this work.}
\affiliation{Department of Applied Physics, Eindhoven University of Technology, 5600 MB Eindhoven, The Netherlands}
\author{Orson A.H. van der Molen}\thanks{These authors contributed equally to this work.}
\affiliation{Department of Applied Physics, Eindhoven University of Technology, 5600 MB Eindhoven, The Netherlands}
\author{Wouter H.J. Peeters}
\affiliation{Department of Applied Physics, Eindhoven University of Technology, 5600 MB Eindhoven, The Netherlands}
\author{Marcel A. Verheijen}
\affiliation{Department of Applied Physics, Eindhoven University of Technology, 5600 MB Eindhoven, The Netherlands}
\affiliation{Eurofins Materials Science Netherlands BV, 5656 AE Eindhoven, The Netherlands}
\author{Erik P.A.M. Bakkers}
\email{e.p.a.m.bakkers@tue.nl}
\affiliation{Department of Applied Physics, Eindhoven University of Technology, 5600 MB Eindhoven, The Netherlands}

\begin{abstract}
Heteroepitaxy enables the engineering of novel properties, which do not exist in a single material.
Two principle growth modes are identified for material combinations with large lattice mismatch, Volmer-Weber and Stranski-Krastanov.
Both lead to the formation of three-dimensional islands, hampering the growth of flat defect-free thin films.
This limits the number of viable material combinations.
Here, we report a distinct growth mode found in molecular beam epitaxy of PbTe on InP initiated by pre-growth surface treatments.
Early nucleation forms islands analogous to the Volmer-Weber growth mode, but film closure exhibits a flat surface with atomic terracing.
Remarkably, despite multiple distinct crystal orientations found in the initial islands, the final film is single-crystalline.
This is possible due to a reorientation process occurring during island coalescence, facilitating high quality heteroepitaxy despite the large lattice mismatch, difference in crystal structures and diverging thermal expansion coefficients of PbTe and InP.
This growth mode offers a new strategy for the heteroepitaxy of dissimilar materials and expands the realm of possible material combinations.
\end{abstract}

\maketitle

\section*{I. Introduction}

Heteroepitaxy has been a staple of modern material science enabling a wide variety of techniques such as band alignment tuning and surface passivation,\cite{Lauhon.2002} crystal structure transfer,\cite{Fadaly.2020} superlattices,\cite{Mundy.2016} strain engineering,\cite{Schaeffler.1997} and virtual substrates.\cite{Bioud.2019}
Strictly two-dimensional layer-by-layer growth ensues if adatoms are more strongly bound to the substrate than to each other.\cite{Frank.1949} For the heteroepitaxy of dissimilar materials, this is generally not possible.\cite{Palmstrom.1995} 
Instead, if the adatoms are more strongly bound to each other than to the substrate, they follow the Volmer-Weber growth mode with the formation of three-dimensional islands.\cite{Volmer.1926} 
Alternatively, in the intermediate Stranski-Krastanov case, initial layer-by-layer growth occurs until a critical thickness is reached where island growth ensues.\cite{Stranski.1937} 
The two latter options lead to a three-dimensional surface topography with a high defect density.\cite{Floro.2001,Mo.1990,Teichert.2002}
This can be detrimental to the desired material characteristics or geometry, limiting the viable material combinations.

A prime example for a research field dependent on artificially structured materials with stringent quality requirements is topological quantum computation.
Here, inherently fault-tolerant qubits have been proposed, based on the non-abelian braiding statistics exhibited by Majorana bound states.\cite{Kitaev.2001,Kitaev.2003,Nayak.2008,Sarma.2015}
Suitable solid-state systems rely heavily on deliberate material design, with proposals suggesting the use of semiconductor nanowire networks on an electrically isolating substrate, partially coupled to epitaxially grown superconducting islands.\cite{Lutchyn.2010,Oreg.2010,Karzig.2017,Plugge.2017}
Despite significant advances in the fabrication of the heterostructures,\cite{Chang.2015,Guel.2017,Krizek.2018,Heedt.2021,Kanne.2021} a definite proof of the existence of Majorana bound states is lacking.
A major challenge is posed by material limitations causing disorder, e.g. surface roughness, charge impurities, point defects, atomic vacancies, patterning imperfections, or geometric restrictions.\cite{Pan.2020,Sarma.2021,Woods.2021,Ahn.2021}
The ability to reduce this disorder is critical for the development of Majorana qubits and solid state based quantum technologies more broadly, making high quality heteroepitaxy imperative.

In this work, we explore the molecular beam epitaxy of PbTe on InP (111)A substrates. The lead-salt is an attractive material choice for topological quantum computation,\cite{Springholz.1993,Grabecki.1999,Grabecki.2004,Chitta.2005,Chitta.2006,Grabecki.2006,Wuttig.2018,Geng.2021,Schellingerhout.2022,Cao.2022,Jiang.2022,Kate.2022,Jung.2022} suppressing disorder due to the screening of charged impurity scattering, resulting from the large dielectric constant.\cite{Yuan.1997,Grabecki.2005} InP is a suitable substrate due to the insulating properties, availability, and well developed processing schemes. 
The growth initially follows the Volmer-Weber model, forming islands that subsequently coalesce, percolate, and finally, in response to a pre-growth surface treatment, form a closed film exhibiting a terrace-stepped surface.
An involved crystal reorientation process facilitates the growth of large single-crystalline PbTe films regardless of the significant lattice mismatch, different crystal structure, and diverging thermal expansion coefficient between growth and substrate. Reorientation processes have previously only been shown in metals compensating small angle mismatches between islands of about~$1$°.\cite{Pashley.1964}
High quality growth on a comparable material combination has been reported, however no reorientation process was observed, as initial islands exhibited only one epitaxial orientation upon surface treatments.\cite{Haidet.2020}
Understanding and exploiting the described growth mechanism can open paths to new high quality heterostructures involving dissimilar materials.

\begin{figure*}[t!]
    \centering
    \includegraphics[width=1\textwidth]{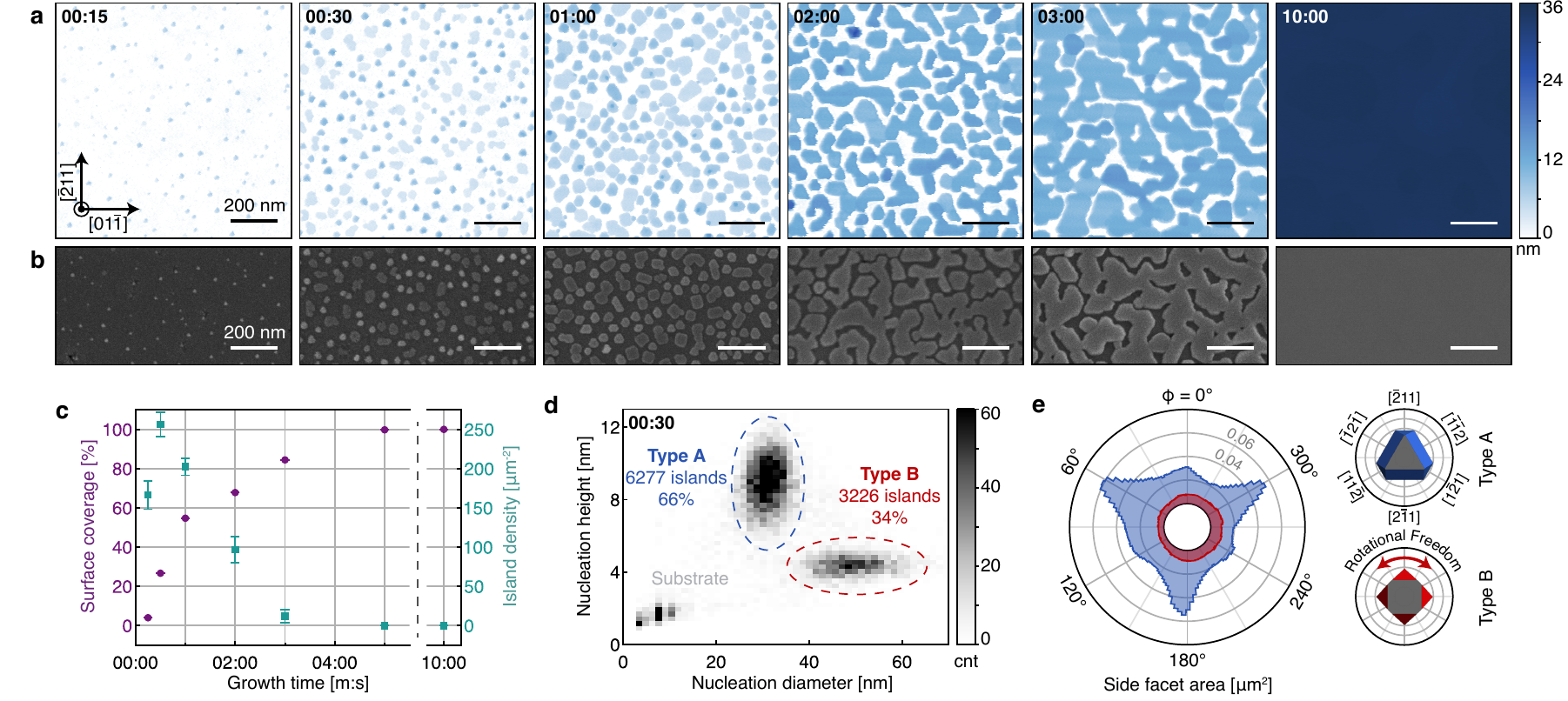}
    \caption{\textbf{PbTe layer formation. a,} AFM scans depict the growth stages from initial island formation, to their coalescence, and the development of a closed layer. The in-plane crystal directions of the InP substrate are indicated in the first panel and kept consistent throughout.
    \textbf{b,} SEM micrographs of the same samples.
    \textbf{c,} Island density (cyan) and surface coverage (purple) plotted over growth time, both extracted from AFM data. The number of islands reaches a maximum at 30~s, after which the probability to form new nuclei decreases and islands begin to coalesce. At 3~min the film has percolated and is almost fully connected. The error bars show the standard deviation across 36 adjacent $1\times1~\upmu$m areas.
    \textbf{d,} A two-dimensional histogram compares diameter and height of islands taken from the AFM data at 30~s growth time. Two distinct types are discernible.
    \textbf{e,} A polar histogram of the side facet area in dependence of the azimuthal angle $\phi$ taken from nuclei in the 30~s AFM data. A cutoff threshold is set for surfaces at polar angles $\theta \leq 20$° relative to the substrate normal. The data indicates a preferential orientation of type A nuclei. This is not observed in type~$B$ nuclei, pointing towards a weak adhesive force between substrate and island. A Wulff construction of the two types suggests the involved facet directions, corresponding to peaks in the histogram.}
    \label{fig1}
\end{figure*}

\section*{II. PbTe Layer Formation}
A time series, depicted in Fig.~\ref{fig1}a-b, explores the growth behaviour of PbTe on (111)A InP substrates. Initially, discrete islands are formed, visible already at 15~s growth time. The islands subsequently expand both vertically and laterally, until they begin to coalesce. The film has almost fully percolated at 3~min, after which it forms a closed layer exhibiting atomic terracing which only forms in consequence of a pre-growth surface treatment (see Supplementary Information Fig.~S\EFfilms).
This film formation behaviour is summarised in Fig.~\ref{fig1}c, where island density and surface coverage are plotted over time.
Initially, the observed increase of surface coverage is driven both by creation of new and expansion of existing islands. However, around 30~s, new nuclei stop forming and islands begin to coalesce, resulting in a decrease in the island density. 
The distributions of island height and diameter support the cessation of new island formation. This is shown for 30~s growth time as a two-dimensional histogram in Fig.~\ref{fig1}d. The lack of a tail into the small island heights and diameters imply a homogeneity in the age of the nuclei, where the presence of a certain density of islands blocks the formation of new islands. This indicates a reasonable diffusion length of growth species over the InP substrate, as atoms impinging on the substrate are prevented from forming a new island by diffusing to an existing one. Consequently, these islands compete for material. Fig.~\ref{fig1}d reveals not only a narrow spread for both displayed quantities, but also a bimodal distribution. This is a result of two distinct types of islands with different aspect ratios, hereafter referred to as type~$A$~(66\%) and type~$B$~(34\%). Analysis across different growth times confirms that islands of both types form before 15~s growth time and grow in both vertical and lateral directions with type dependent rates (see Supplementary Information Fig.~S\EFhistograms).
Using surface normals extracted from the atomic force microscopy (AFM) data, the polar plot in Fig.~\ref{fig1}e reveals the distribution of side facet orientations of the 30~s grown islands, separated by type. The presence of maxima indicates a preferential epitaxial orientation of Type~$A$ islands, with the corresponding facets indicated in the inset Wulff-construction. The absence of any preferred orientation in type~$B$ islands suggests by contrast an in-plane rotational freedom.

\begin{figure*}[t!]
    \centering
    \includegraphics[width=1\textwidth]{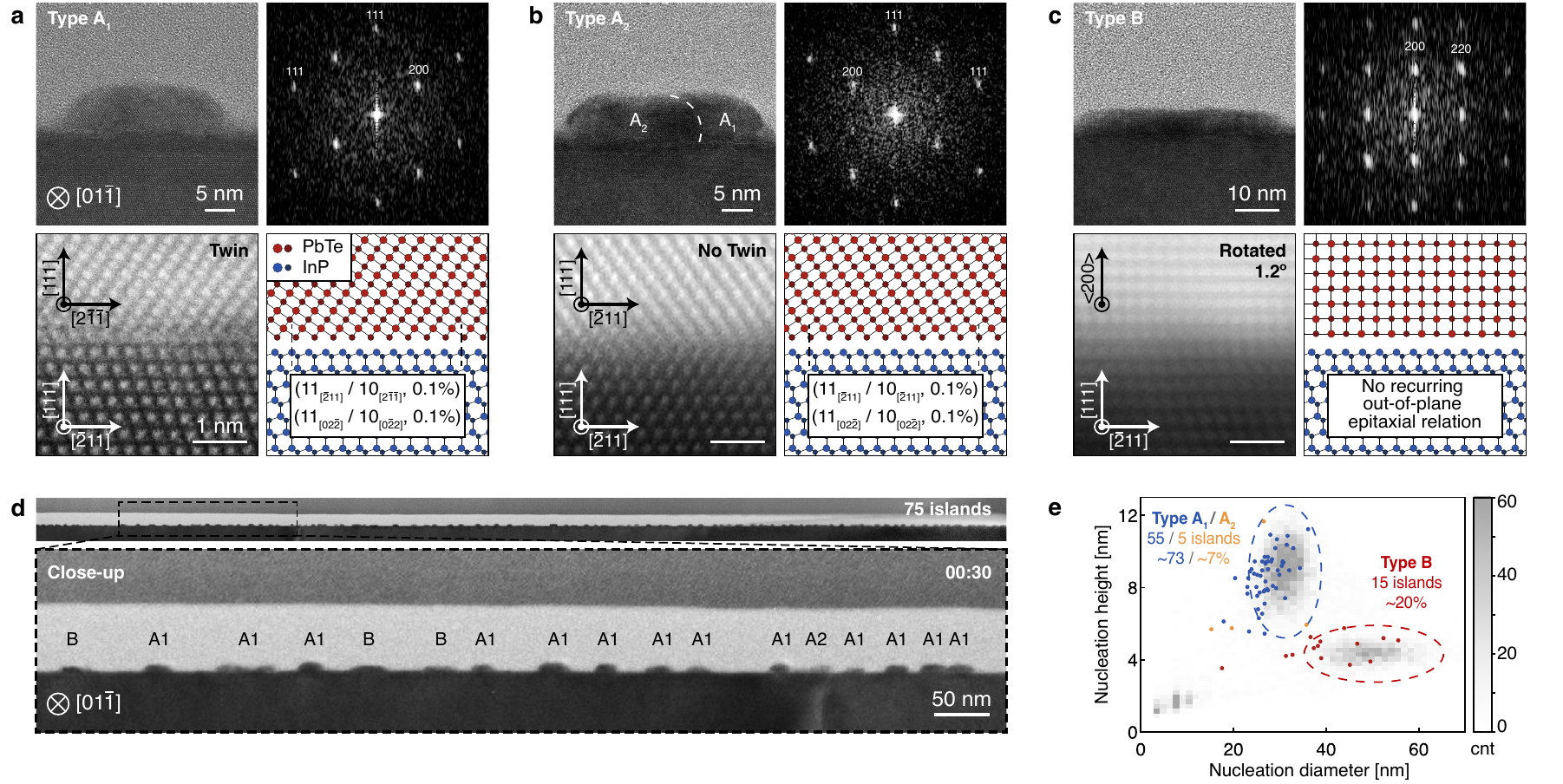}
    \caption{\textbf{Epitaxial orientation of islands. a-c,} Islands exhibit three types of epitaxial orientation to the substrate. For each type are shown, a representative TEM micrograph, the two-dimensional FFT of aforementioned micrograph, a HAADF scanning TEM micrograph of an equivalent interface, and the corresponding structural model of the crystal lattice. The TEM micrograph of Fig.~\ref{fig2}b shows two superimposed islands, each with their distinct type indicated.
    \textbf{d,} Bright-field TEM micrograph of a cross-section taken from the 30~s growth time sample shown in Fig.~1a. A close-up displays islands and their type.
    \textbf{e,} Height and diameter of the islands measured with TEM. Data is in line with the the histogram from Fig.~\ref{fig1}d, which confirms the connection between epitaxial orientation and the island types found from AFM.}
    \label{fig2}
\end{figure*}

\section*{III. Epitaxial Orientation of Islands}

Transmission electron microscopy (TEM) analysis of the 30~s grown sample reveals three island types, each defined by a distinct epitaxial relation to the substrate. A representative high-resolution TEM image of the most frequent type, $A_1$, is shown in Fig.~\ref{fig2}a. This type is characterised by a twinned epitaxial relation between PbTe and InP, as confirmed by the micrograph's fast Fourier transform (FFT). This results in a $[\bar{2}11]_{\text{InP},\perp}/[2\bar{1}\bar{1}]_{\text{PbTe},\perp}$ and $[02\bar{2}]_{\text{InP},\parallel}/[0\bar{2}2]_{\text{PbTe},\parallel}$ interface along the in-plane directions transverse and parallel to the depicted zone axis. A high-angle annular dark-field (HAADF) scanning TEM image of an equivalent interface reveals the atomic planes of both crystal phases and establishes a structural model of the lattice. The large lattice mismatch of $10.1\%$ is overcome through the formation of edge-type misfit dislocations at the InP-PbTe interface, breaking bonds in exchange for a reduction of strain. For type~$A_1$, an $11_{[\bar{2}11]}/10_{[2\bar{1}\bar{1}]}$ lattice plane ratio is found in-plane, leading to 0.1\% residual mismatch in the corresponding ideal flat interface. 
The less frequent type~$A_2$ shown in Fig.~\ref{fig2}b exhibits no interfacial twinning. The $[\bar{2}11]_{\text{InP},\perp}/[\bar{2}11]_{\text{PbTe},\perp}$ and $[02\bar{2}]_{\text{InP},\parallel}/[02\bar{2}]_{\text{PbTe},\parallel}$ crystal directions are a direct continuation of the substrate, with an identical lattice plain ratio and residual mismatch as type~$A_1$ due to the structural similarity.
In contrast to both types~$A_i$, type~$B$ changes out-of-plane crystal direction at the interface from [111] to $\langle200\rangle$. Despite this distinct out-of-plane direction, no preferential in-plane orientation can be found, suggesting a weak adhesive force between the substrate and this island type.\cite{Miceli.1995} Due to this, no recurring lattice plain ratio or residual mismatch can be assigned to this island type.
A bright-field TEM micrograph of the complete cross-section with 75~islands is shown in Fig.~\ref{fig2}d. Several labelled islands can be seen in the close-up. In Fig.~\ref{fig2}e the diameter and height values of islands found in the cross-section cut, measured via TEM, are compared to the AFM data presented in Fig.~\ref{fig1}e. Tip convolution effects during the AFM measurement likely lead to an overestimation of the island diameter, in addition to the uncertainties introduced by the TEM projection. Despite these inherent inaccuracies, a close agreement between these measurement modalities is found, connecting the TEM based epitaxial relation of the island types with the superior statistics of the AFM data. 

\begin{figure*}[t!]
    \centering
    \includegraphics[width=1\textwidth]{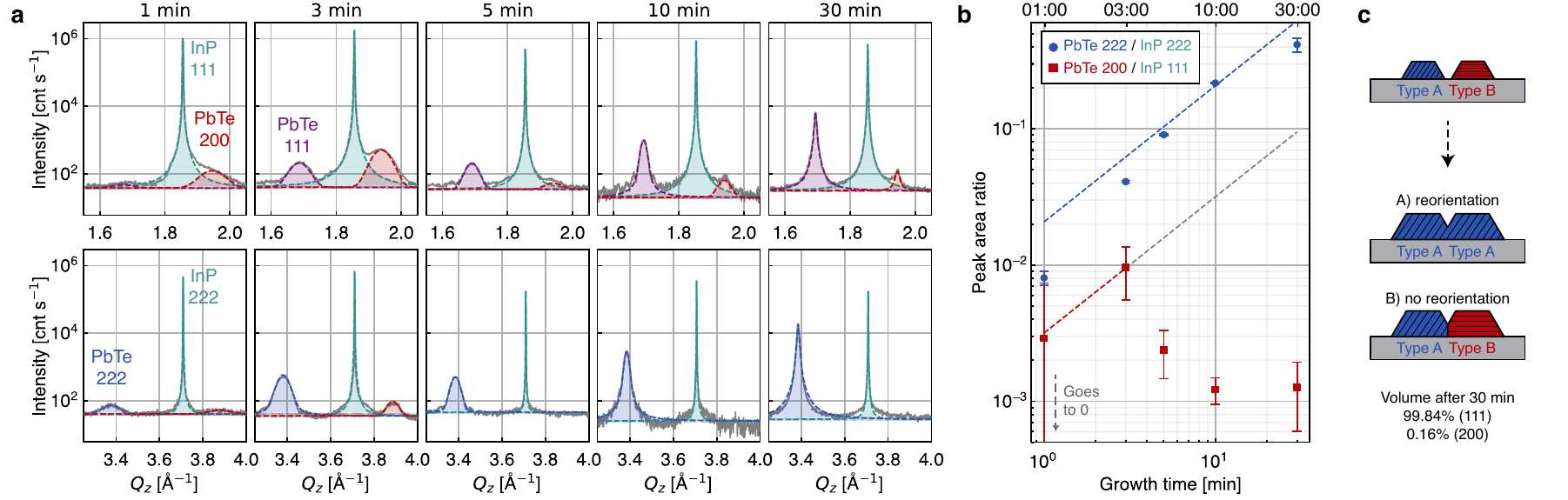}
    \caption{\textbf{Reorientation upon island coalescence. a,} Symmetric $\upomega$-$2\uptheta$ XRD scans of PbTe grown on InP substrates. The 1st (top row) and 2nd (bottom row) order peaks are plotted for increasing growth times.
    \textbf{b,} The peak area ratio between PbTe and InP is indicative of the probed crystal volume.
    Opposed to the continuously increasing (111)-oriented PbTe crystal volume, the (200)-oriented volume decreases by nearly an order of magnitude following 3~min growth time. This is suggestive of a crystal reorientation process of islands with diverging epitaxial relation, triggered upon coalescence. The remaining (200) signal at 30~min growth time implies that nearly all undergo this process, with (0.16±0.06)\% of the film volume estimated to remain (200)-oriented (see Supplementary Information Fig.~S\EFfullxrd).
    \textbf{c,} A schematic illustrating the suggested reorientation process. Upon coalescence, type~$B$ islands can adapt their crystallographic orientation.}
    \label{fig3}
\end{figure*}

\section*{IV. Reorientation Upon Coalescence}

In light of the various crystal orientations present during the initial growth stages, it is of particular interest to study the mechanics leading to a closed film with atomically flat terraces. The stable lattice plane orientation throughout the growth (see Supplementary Information Fig.~S\EFplaneorientation) enables their study through symmetric $\upomega$-$2\uptheta$ XRD scans. Fig.~\ref{fig3}a plots sections of the scattered X-ray intensity for increasing growth times. Fitting the XRD-spectra with pseudo-Voigt functions allows for the identification of the isolated InP~(111), PbTe~(111), and PbTe~(200) peaks and their higher order reflections. The two PbTe peaks correspond to type~$A_i$ and $B$ epitaxy, respectively. No indications of other orientations can be found, confirming the absence of additional types (see Supplementary Information Fig.~S\EFfullxrd). 
The area under each peak is correlated with the probed volume of that crystal orientation.\cite{Als.2020} Fig.~\ref{fig3}b compares the peak area of PbTe~(222) and PbTe~(200) with growth time as an indication of the crystal growth evolution. Each calculated area is normalised with the InP peak area to eliminate any influence originating from varying sample size and alignment. 
The first noticeable characteristic of the plot is the steady increase of the PbTe~(222) volume. As expected, the same behaviour is found for PbTe~(111) (see Supplementary Information Fig.~S\EFfullxrd). In contrast to this, following an initial increase, the probed (200)~crystal volume decreases by nearly an order of magnitude. No new orientations appear in the XRD~spectra, pointing towards a reorientation process of the initial type~$B$ nuclei into type~$A_i$. Based on the growth time dependency, the reorientation likely takes place upon island coalescence up until the subsequent film percolation. A sketch of the proposed process is shown in Fig.~\ref{fig3}c. 
However, not all islands undergo this reorientation process as evident from the remaining PbTe~(200) signal at 30~min growth time. This is supported by the continuously decreasing (200)~peak width in Fig.~\ref{fig3}a, a sign of ongoing vertical growth of those grains. Together with the initial (200)~area decrease, this excludes overgrowth of the nuclei as a possible explanation of the observed phenomena, and instead suggests the coexistence of a small volume fraction of $(0.16\pm0.06)\%$ remaining (200)~type~$B$ grains at the PbTe growth front (see Supplementary Information Fig.~S\EFfullxrd). We note that this can likely be further optimised. 
Reciprocal space maps of asymmetric reflections distinguish twinned and non-twinned (111)~PbTe layers, and are used to quantify the ratio between type~$A_1$ and type~$A_2$ in the layer (see Supplementary Information Fig.~S\EFtwinning). Already at 3~min growth time all (111)~oriented PbTe is found to be twinned relative to the InP substrate, i.e. only type~$A_1$ remains. This is consistent with respective TEM observations.\cite{Jung.2022}
Island orientation and morphology are governed by the interfacial and surface free energies. Additionally, there can be a bulk contribution in the form of strain energy, and contributions stemming from defects or grain boundaries. For reorientation to occur, it must both lead to a lower energy state and not have too high of an energy barrier. The relative influence of the substrate on the islands is expected to decline over time as the surface to volume ratio decreases. As such, it is unlikely that the reorientation of islands occurs spontaneously as the islands grow. The introduction of grain boundaries between different island types is therefore identified as the most probable trigger for reorientation. As the reorientation process occurs predominately to exclusively from type~$A_2$ and $B$ towards type~$A_1$, the substrate interface plays a directing role in the reorientation, suggesting a minimal interfacial energy for type~$A_1$ islands.
It is difficult to overemphasise the importance of the reorientation process, as it facilitates the heteroepitaxy of high-quality thin films on dissimilar substrates, in this case PbTe~(111) on InP~(111)A. In fact, the seemingly unfit substrate with its large lattice mismatch, different crystal structure, and diverging thermal expansion coefficient, are believed to lead to a weak adhesive force between substrate and growth, facilitating the reorientation process. 

\begin{figure*}[t!]
    \centering
    \includegraphics[width=1\textwidth]{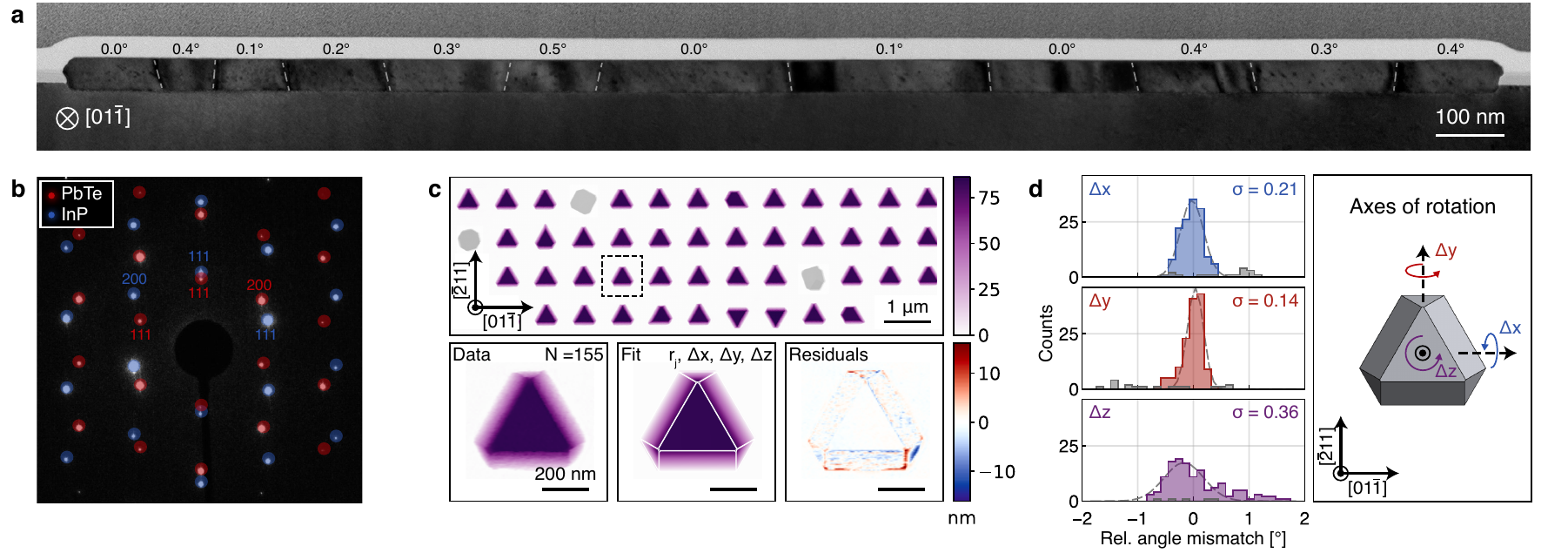}
    \caption{\textbf{Layer mosaicity. a,} Cross-section of a $2\times2~\upmu$m PbTe structure, captured via bright-field TEM. The image is taken off zone axis to emphasise contrast variations throughout the film originating from strain between merged segments (separated by dashed lines). The misalignment, i.e. the relative rotation around the substrate normal, between each segment and the substrate is indicated. \textbf{b,} Despite this, the complete PbTe segment is single-crystalline, as demonstrated by a representative diffraction pattern taken from a full data set (see Supplementary Information Fig.~S\EFtem). \textbf{c,} AFM scan of SAG structures grown from circular mask openings, showing well-defined $\{200\}$ and $\{111\}$ facets. Facet radii and three-axis rotation are fitted to each structure, with a typical structure, corresponding fit, and residual shown in the bottom panels. \textbf{d,} Distribution of rotations for each axis across 155 structures. Bad fits with a relative residual volume above 4\% are plotted in grey. The axis definitions are shown in the right panel.}
    \label{fig4}
\end{figure*}

\section*{V. Layer Mosaicity}

The crystal quality resulting from the reorientation process is assessed via TEM. A cross-section of a selective area grown~(SAG) $2\times2~\upmu$m structure shows subtle contrast variations throughout the film. These become pronounced when imaged off zone axis, as depicted in Fig.~\ref{fig4}a.
Despite these boundaries and despite the film originating from many islands with various epitaxial orientations, the complete cross-section has a single twinned epitaxial relation to the substrate, confirming the observations made in Fig.~\ref{fig3}. This finding is supported by equivalent electron diffraction patterns taken across the complete segment (see Supplementary Information Fig.~S\EFtem). A representative example is shown in Fig.~\ref{fig4}b.
The contrast variations visible in Fig.~\ref{fig4}a are identified as boundaries between slightly misoriented segments, where the misorientation is a rotation around the surface normal of the substrate. The rotations are distributed between 0° and 0.5° and have a positive mean, indicating that in addition to individual variation, there is a common rotation relative to the substrate. The average relative rotation between neighbouring segments is 0.2°. This corresponds to a shift of one lattice plane about every 287 columns (about 185~nm in the PbTe crystal), and therefore amounts to, at most, a few planes offset over the length of each segment. This confirms that the PbTe films are single-crystalline, accompanied by slight mosaicity.
The segment boundaries are not a result of incomplete reorientation between coalescing pairs of islands of different types. Based on the AFM data presented in Fig.~\ref{fig1}, on average 7 islands, of which 2 type~$B$, combine to form a single segment of type~$A_1$. This suggests that multiple islands combine and reorient to form a strain free segment.
To inspect the origin of the strain features, nanostructures were selectively grown from circular openings with a 200~nm diameter as shown in Fig.~\ref{fig4}c. Based on the segment dimensions visible in TEM, the structures are expected to consist on average of a single segment. As such, measurements of their orientations can be compared to segment orientations in films before their merging. The facets formed in the nanostructures belong to the $\{111\}$ and $\{200\}$ families and can be reproduced via Wulff construction.\cite{Jung.2022} Based on the crystal symmetry, a fitting function is defined and used to acquire both facet radii and misorientation in three axes of rotation. A representative fit is shown in the bottom panel of Fig.~\ref{fig4}c. A histogram over 155 fitted structures is depicted for each rotational axis in Fig.~\ref{fig4}d. The found standard deviation of the out of plane rotation $\mathrm{\Delta z}$ is 0.36°. This includes noise from imperfect facets, AFM measurement, and fitting and therefore gives an upper bound to the variation in structure orientations. The distribution suggests, that the origin of the strain features can be found in slight misalignments between meeting grains, that have become too large in size to align completely.

\section*{VI. Conclusion}

To summarise, we present a distinct growth mode facilitating high quality heteroepitaxy of dissimilar materials by example of PbTe on InP (111)A substrates. AFM measurements reveal the three-dimensional islands present in early stages of the growth, and the eventual formation of a closed film. Structural differences in the initial islands are attributed to three distinct epitaxial orientations. XRD scans following the subsequent growth stages expose a reorientation process, facilitating the formation of a predominantly single-crystalline film. A fascinating outcome, considering the initial differences in epitaxial orientation of the nuclei. Small-angle misalignment is detected between segments of the film leading to strain signatures exposed by off zone axis TEM. Nevertheless, our recent quantum experiments on selective area grown structures employing the same growth mechanism yield electron mobilities comparable to InSb and a coherence length exceeding any previously reported values on selective area grown networks, signifying the high crystal quality of the PbTe.\cite{Jung.2022} Future work will explore the heteroepitaxial growth of topological crystalline insulators, namely SnTe and PbSnTe, that are expected to exhibit comparable growth behaviour.

\section*{Methods}
\noindent\textbf{Substrate fabrication}
Undoped semi-insulating (111)A InP substrates were used as growth substrates. Selective area growth required additional processing as described in an accompanying publication.\cite{Jung.2022}\\

\noindent\textbf{PbTe heteroepitaxy}
Growth took place in an ultra-high vacuum molecular beam epitaxy system. An annealing step under Te overpressure at 480~°C was used for surface reconstruction of the etched openings and to remove oxide residuals from the exposed substrate surface. The PbTe films were subsequently grown at 340~°C with separate elemental sources, providing a Te flux of about $4.00\times10^{-7}$~mbar and a Pb flux of about $1.25\times10^{-7}$~mbar measured as beam equivalent pressure using a naked bayard-alpert ion gauge. Temperatures were measured with a kSA BandiT system based on the optical absorption edge.\\

\noindent\textbf{TEM studies}
TEM lamellas were prepared in a FEI Nova Nanolab~600i as described in a previous publication.\cite{Jung.2021} TEM studies were performed using a probe-corrected JEOL~ARM~20OF, equipped with a 100~mm$^2$ Centurio SDD Energy dispersive X-ray spectroscopy detector. 
Reported grain misorientations in Fig.~4a are based on tilt differences between the zone axis of the substrate and each individual grain in scanning TEM.\\

\noindent\textbf{Modelling}
All structural simulations of the NW crystals in Figs~2 were done using the Vesta software.\cite{Momma.2011}\\

\noindent\textbf{X-ray crystallography}
XRD studies were carried out using a PanAlytical X'Pert Pro MRD diffractometer, equipped with a Cu K-$\upalpha$ radiation source. The scans were taken with a 1D~detector, with pixels distributed along the 2$\uptheta$ direction. The detector width was set to 1.2° degrees to suppress the background for the 1~min growth sample. Other measurements used the full detector width of 2.5°.
Large-scale statistics about the evolution of PbTe~(200) and PbTe~(111) were obtained using a beam spot of approximately $1\times3$~mm. Fitting the spectra was done by fixing the background level on the average number of counts between $Q_z = 2.2-2.8$~1/Å.
The normalisation with the InP substrate peak implicitly assumes that the PbTe film is transparent for X-rays. This is not completely correct, as the heavy elements Pb and Te are efficient at scattering X-rays. As a result, normalisation with the InP peak overestimates the PbTe/InP ratio. This overestimation becomes worse with increasing film thickness, and is therefore not an explanation for the observed (200) decrease after 3~mins of growth.\\

\noindent\textbf{AFM studies and data processing}
Atomic force microscopy was performed using a Bruker Dimension Icon and a Bruker SCANASYST-AIR probe with a 2 nm tip radius and 0.4 N/m spring constant (nominal) via the SCANASYST acquisition mode. The scan size was $6.4\times6.4~\mathrm{\upmu m}$ with a resolution of 2048 by 2048 pixels and a scan rate of 0.5 Hz.

\noindent\textit{Detilting:} The scans were detilted in two steps. First, by subtracting a plane with with an orientation matching the median gradient found in the scan and second by fitting and subtracting a line to each line in the scan direction. 

\noindent\textit{Substrate level calibration:} Zero height was determined for the scans with surface coverages below 100\% by fitting a bimodal Gaussian distribution to a histogram of the entire detilted scan. The lower Gaussian's center height was then defined to be the substrate surface level. For scans with surface coverages of 100\%, a separate scan was carried out on the boundary between the mask and the opening, extracting the height above the mask of the structure and adding that to the thickness of the mask measured on a different sample from the same wafer.

\noindent\textit{Fitting:} 
Fitting in panel~4c is carried out through least-squares minimisation. The in-plane structure's centre is also fitted, this removes the necessity of independently fitting the radii of the three $\{200\}$ facets. A correlation matrix for all fitting parameters is shown in Supplementary Information Fig.~S\EFfits.

\bibliographystyle{naturemag}
\bibliography{ref.bib}

\vspace{.5cm}

\noindent\textbf{Acknowledgements}
We thank NanoLab@TU/e for their help and support.
This work has been supported by the European Research Council (ERC TOCINA 834290) and TOPSQUAD (Grant No. 862046).
We furthermore acknowledge Solliance, a solar energy R\&D initiative of ECN, TNO, Holst, TU/e, IMEC and Forschungszentrum Jülich, and the Dutch province of Noord-Brabant for funding the TEM facility.\\

\noindent\textbf{Author Contributions}
J.J. and S.G.S. carried out the substrate fabrication and the growth of PbTe. 
O.A.H.v.d.M. and J.J. performed the AFM characterisation and data analysis.
W.H.J.P. performed and analysed the XRD measurements.
M.A.V. performed the TEM analysis.
J.J. and S.G.S. prepared FIB lamellae.
M.A.V. and E.P.A.M.B. provided key suggestions and discussions and supervised the project.
J.J. and O.A.H.v.d.M. wrote the manuscript, with contributions from all authors.\\

\noindent\textbf{Competing Interests} 
The authors declare no competing interests.\\

\noindent\textbf{Additional Information}
Supplementary Information is available for this paper.\\
Correspondence should be addressed to to E.P.A.M.B. (e.p.a.m.bakkers@tue.nl).\\

\noindent\textbf{Data Availability}
The data supporting the findings of this study is openly available at https://doi.org/10.5281/zenodo.6900774

\end{document}


\title{Supplementary: Single-crystalline PbTe film growth through reorientation}

\author{Jason Jung}\thanks{These authors contributed equally to this work.}
\affiliation{Department of Applied Physics, Eindhoven University of Technology, 5600 MB Eindhoven, The Netherlands}
\author{Sander G. Schellingerhout}\thanks{These authors contributed equally to this work.}
\affiliation{Department of Applied Physics, Eindhoven University of Technology, 5600 MB Eindhoven, The Netherlands}
\author{Orson A.H. van der Molen}\thanks{These authors contributed equally to this work.}
\affiliation{Department of Applied Physics, Eindhoven University of Technology, 5600 MB Eindhoven, The Netherlands}
\author{Wouter H.J. Peeters}
\affiliation{Department of Applied Physics, Eindhoven University of Technology, 5600 MB Eindhoven, The Netherlands}
\author{Marcel A. Verheijen}
\affiliation{Department of Applied Physics, Eindhoven University of Technology, 5600 MB Eindhoven, The Netherlands}
\affiliation{Eurofins Materials Science Netherlands BV, 5656 AE Eindhoven, The Netherlands}
\author{Erik P.A.M. Bakkers}
\email{e.p.a.m.bakkers@tue.nl}
\affiliation{Department of Applied Physics, Eindhoven University of Technology, 5600 MB Eindhoven, The Netherlands}

{\let\clearpage\relax\maketitle}

\section*{Volume fraction calculation}

A quantitative description of the reorientation process is made by estimating the probed crystal volumes, i.e. the number of unit cells contributing to the signal. From experimental data the integrated intensity of the X-ray diffraction peaks~$I_i$ is known, with the subscript~$i$ indicating the reflection of interest. In the kinematical scattering approximation,\cite{Als.2020} $I_i$ is related to the number of unit cells~$N_i$ according to

\begin{equation}
I_i=\Phi_0\:\frac{\lambda^3}{\nu_c} r_0^2 \:\left| S_{G_i}\right|^2 P(\theta_i)\:N_i.
\end{equation}

Here, $\Phi_0$ is the incidence flux of X-rays, $\lambda$ the x-ray wavelength, $\nu_c$ the volume of the unit cell, $r_0$ the Thomson scattering length, and $\left|S_{G_i}\right|$ the magnitude of the structure factor of reflection $i$. The parameter $P(\theta_i)$ contains three terms dependent on the incidence angle $\theta_i$,

\begin{equation}
P(\theta_i)=\frac{1}{\sin{\left(2\theta_i\right)}}\: \frac{1}{\sin{(\theta_i)}}\:\left(1+\cos^2{(2\theta_i)}\right). 
\end{equation}

The term $1/\sin(2\theta_i)$ is known as the Lorentz factor, $1/\sin(\theta_i)$ corrects for the probed sample area, and $\left(1+\cos^2{\left(2\theta_i\right)}\right)$ corrects for the unpolarised source used in the experiment.
The (111)-oriented PbTe is slightly strained (see Supplementary Information Fig.~S\EFtwinning), but the change in unit cell volume compared to (200)-oriented PbTe is neglected. As a result, $\Phi_0$, $\lambda$, $\nu_c$ and $r_0$ are constants independent of the reflection measured. Under these conditions, eq.~(1) can be rewritten to express the volume fraction between reflection $i$ and $j$ as

\begin{equation}
\frac{N_i}{N_i+N_j} = \frac{I_i \: P(2\theta_j) \left|S_{G_j}\right|^2}{I_i \: P\left(2\theta_j\right) \left|S_{G_j}\right|^2 + {I_j \: P(2\theta_i) \left|S_{G_i}\right|}^2}.
\end{equation}

An estimation of the probed crystal fraction requires therefore the evaluation of $P(\theta_i)$, based on eq.~(2), and $\left|S_{G_i}\right|$. For both the~(111) and the~(200) orientation, a relaxed PbTe crystal is assumed with a lattice constant of 6.46~Å. The resulting structure factors, calculated with the python package \textit{xrayutilities}, are summarised in table~S1.

\begin{table}[H]
	\centering
	\caption{\textbf{Structure factors.} Estimated $\left|S_{G_i}\right|$ for relevant InP and PbTe diffraction peaks based on $2\theta_i$ and $P(\theta_i)$.\vspace{.1cm}}
	\begin{tabular}{lc|ccc}
		\toprule
		  & $i$ & $2\theta_i~[\degree]$ & $P(\theta_i)$ & $\left|S_{G_i}\right|$ \\
		\colrule
		InP & 111~   & 26.28 & 17.92 & 182.83 \\
         & 222~   & 54.09 & 3.65 & 103.44 \\
         \colrule
		PbTe & 111~  & 23.87 & 21.94 & 100.86 \\
		& 222~  & 49.04 & 4.56 & 395.14 \\
         & 200~  & 27.35 & 16.47 & 458.14 \\
         & 400~  & 56.80 & 3.26 & 374.57 \\
		\botrule
	\end{tabular}
\end{table}

The PbTe (200)~volume fraction over growth time is plotted in Supplementary Information Fig.~S5c. The fraction at 1~min is estimated around 25-42\%, with a large uncertainty due to the low volume. This coincides with a calculated volume fraction of (38.02$\pm$0.08)\% type~$B$ based on the AFM data shown in Fig.~1d for 30~s growth time. After 30~mins of growth only (0.16±0.06)\% of the film remains in the (200) orientation.

\clearpage
\begin{figure}
	\centering
	\includegraphics[width=\textwidth]{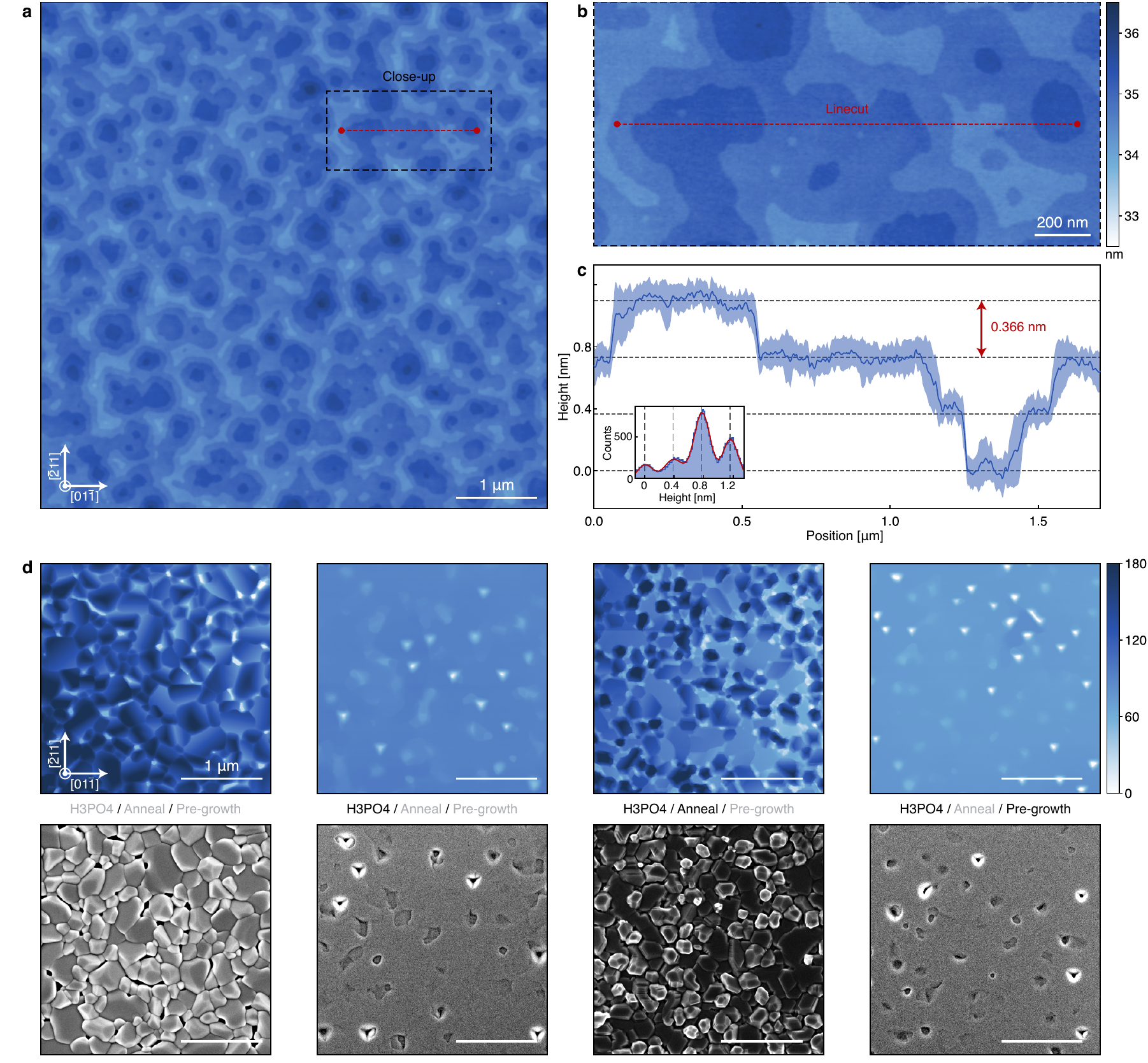}
	\caption{\textbf{Surface terracing and treatment. a,} AFM measurements of a surface after 10~min growth time display a surface characteristic of two-dimensional nucleation following a birth-and-spread model of radially expanding mono-layer steps. The surface roughness is 0.25~nm, based on the arithmetic average of profile height deviations from the mean line. There are five~atomic steps between the highest and the lowest plane. \textbf{b,} Close-up of the region from which a linecut is taken. \textbf{c,} A linecut, averaged over 44~lines spanning a width of 140~nm, with a length of 1.7~$\mathrm{\upmu m}$. The inset shows a histogram of the height data. From this a mono-layer step height of 0.366~nm was estimated (dashed lines), which is in good agreement with the 0.373 nm Pb(111) interplanar spacing. The linecut region was detilted separately and offset by 34.4~nm to align zero to the lowest atomic plane. \textbf{d,} Comparison of the effects of of the surface treatments used for all presented growth. The top row shows AFM scans and the bottom SEM micrographs of the same sample. The scale bar matches panel~\textbf{a}. The colour scale for the AFM scans shows the full height range, with zero aligned to the substrate level.
	The treatments are indicated as \textit{H3PO4}, \textit{Anneal}, and \textit{Pre-growth}.
	First, a phosphoric acid wet etch (H$_2$O\::\:H$_3$PO$_4$ = 10\::\:1) removes the native substrate oxide, allowing epitaxial growth.
	Remaining oxide residuals cause pits in the layer, and require a subsequent annealing step at 480~°C.
	These elevated temperatures damage however the substrate and lead to a  polycrystalline film.
	To mitigate this damage, the anneal is conducted under Te overpressure. This likely additionally supports the epitaxy through the formation of an interfacial InTe layer, with similar reports found in a comparable material combination.\cite{Haidet.2020}
	The Te treatment alone is not sufficient to reach the desired film quality, and still exhibits pits.
	Only by combining all treatments a fully closed film with atomic terracing is observed. 
	Similar behaviour is found in selective area growth.\cite{Jung.2022}}
\end{figure}

\clearpage
\begin{figure}
	\centering
	\includegraphics[width=\textwidth]{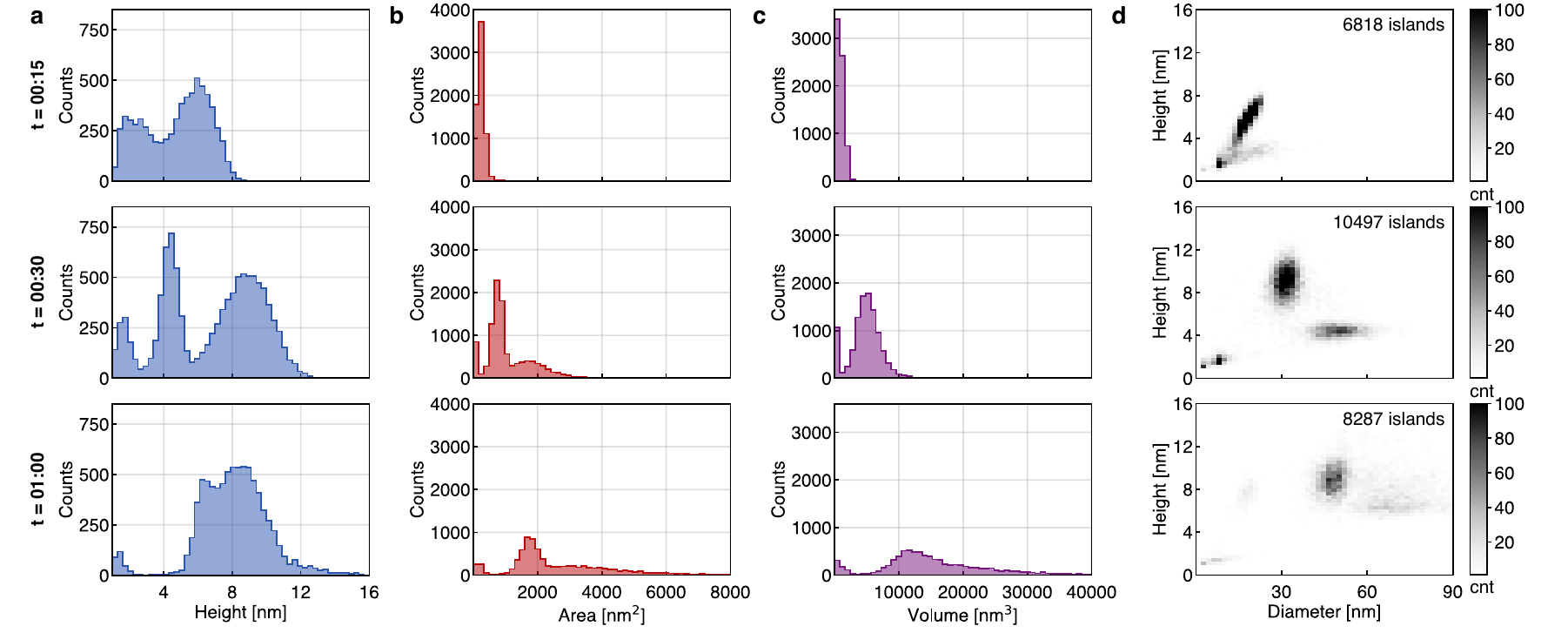}
	\caption{\textbf{Island statistics. a-c,} Histograms of island height, area and volume for multiple growth times. \textbf{d,} Two-dimensional histogram comparing diameter and height of islands. The data at 30~s is also presented in Fig.~1d. All plotted quantities, height, area and volume increase over time, except for the height of type~A islands at 1~minute growth time In the case of height and area, both island types can be distinguished. For volume, this is not apparent, indicating that both types are of similar volume, and only separated in their growth behaviour by a different distribution of material across lateral and vertical growth. The continuous volume increase of both islands types excludes type-selective Ostwald ripening as a dominant growth mechanism.}
\end{figure}

\clearpage
\begin{figure}
	\centering
	\includegraphics[width=\textwidth]{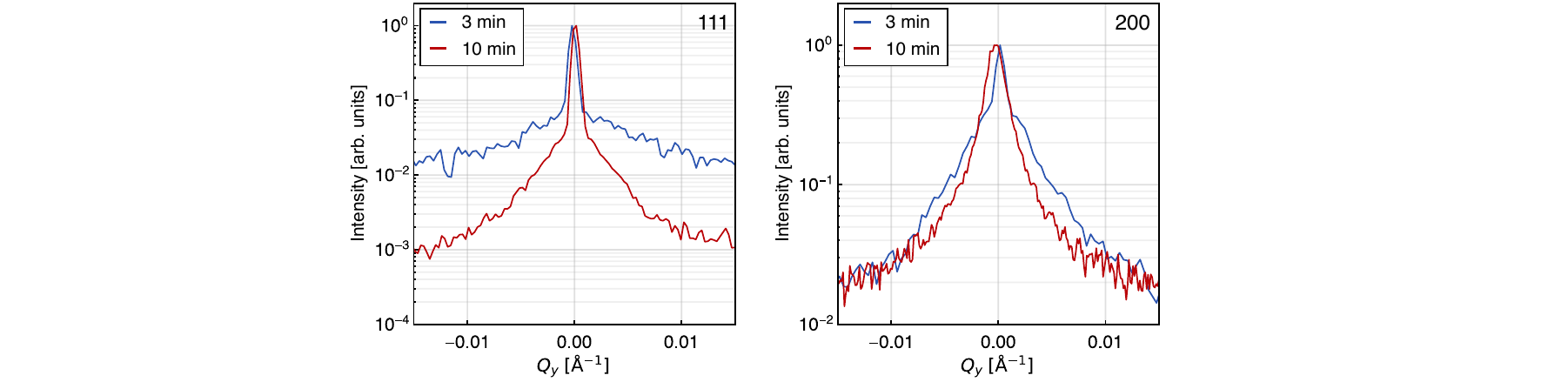}
	\caption{\textbf{Lattice plane orientation stability.} Diffraction peak of the (111) and (200) oriented growth. The lattice plane orientations show no significant changes over growth time allowing their study through symmetric $\upomega$-$2\uptheta$ XRD scans. The data is obtained by integrating a reciprocal space map along $Q_z = 1.693 \pm 0.01$ and $Q_z = 1.947 \pm 0.01$ \text{\AA}$^{-1}$ for the (111) and (200) peaks, respectively.}
\end{figure}

\clearpage
\begin{figure}
	\centering
	\includegraphics[width=\textwidth]{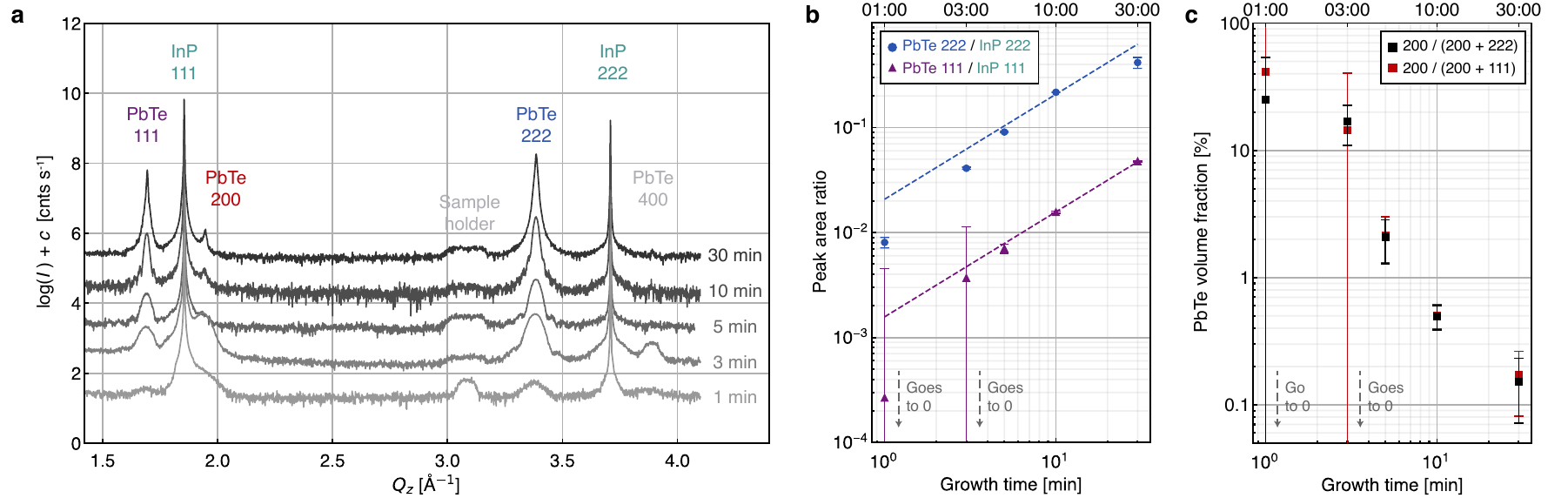}
	\caption{\textbf{XRD orientation data. a,} Continuous spectrum of the symmetric $\upomega$-$2\uptheta$ XRD scans for increasing growth time, offset vertically. Only signatures originating from (111) and (200)~oriented PbTe and the (111)~InP substrate are visible. Features of the sample holder are visible but do not overlap with any diffraction peaks. \textbf{b,} Analogous to the PbTe~(222)/InP~(222) peak area ratio, PbTe~(111)/InP~(111) steadily increases. \textbf{c,} The (200) PbTe peak intensity, and with that the volume fraction, of (200) oriented PbTe continuously decreases. After 30~mins of growth only (0.16±0.06)\% of the film remains in the (200) orientation.}
\end{figure}

\clearpage
\begin{figure}
	\centering
	\includegraphics[width=\textwidth]{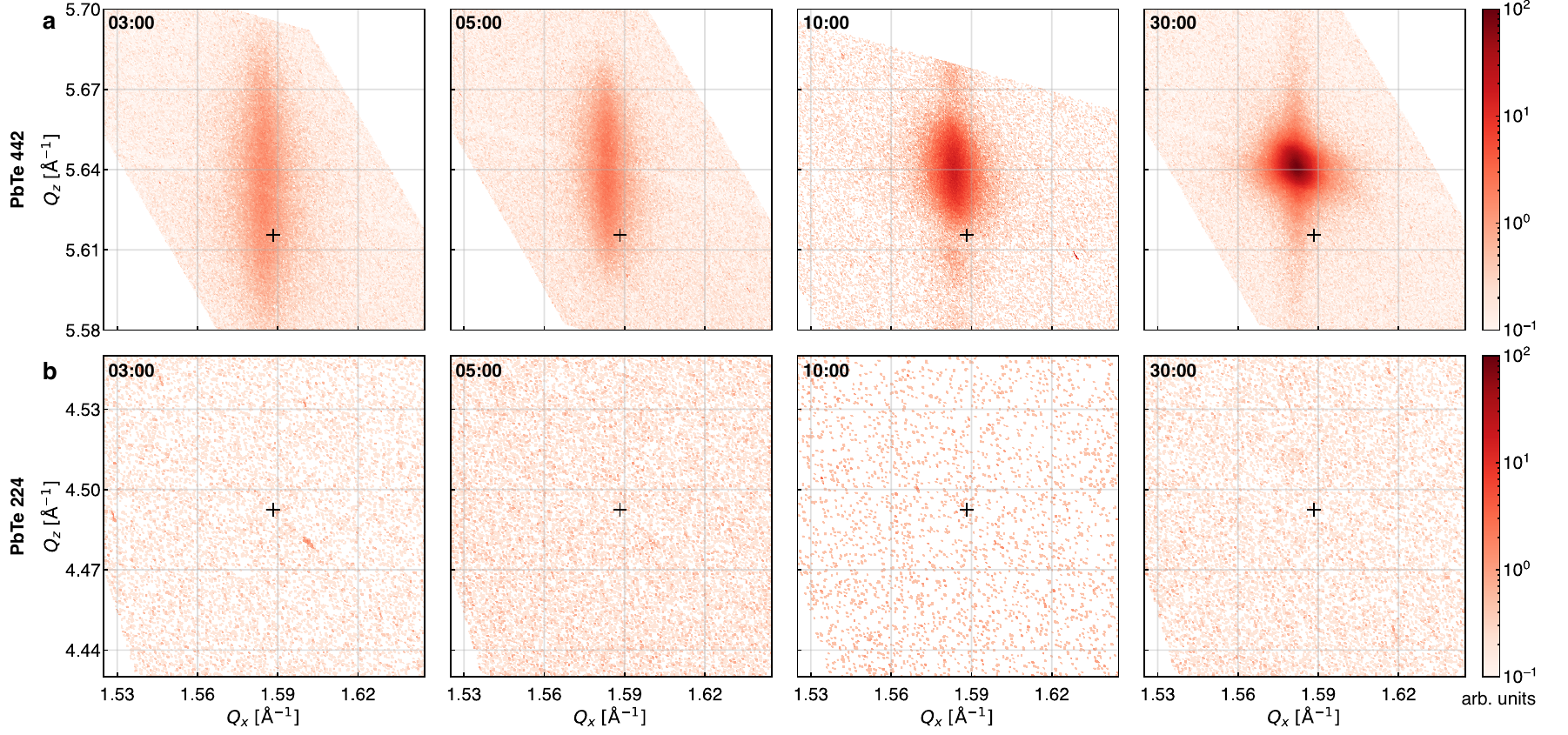}
	\caption{\textbf{Twinning and residual strain. a-b,} Reciprocal space maps of the PbTe~442 and 224 reflections over growth time. 
	The absence of the latter shows that all PbTe (111) is found to be twinned relative to the InP substrate, consistent with TEM observations after film closure (see Supplementary Information Fig.~S\EFtem). 
	The peak narrows with growth time due to the increasing PbTe volume. 
	The offset of the expected PbTe reflection position (black cross) indicates residual strain with in- and out-of-plane lattice parameters converging towards values of $6.49\pm0.01$ and $6.430\pm0.005~ \text{\AA}$, respectively.} 
\end{figure}

\clearpage
\begin{figure*}
	\centering
	\includegraphics[width=\textwidth]{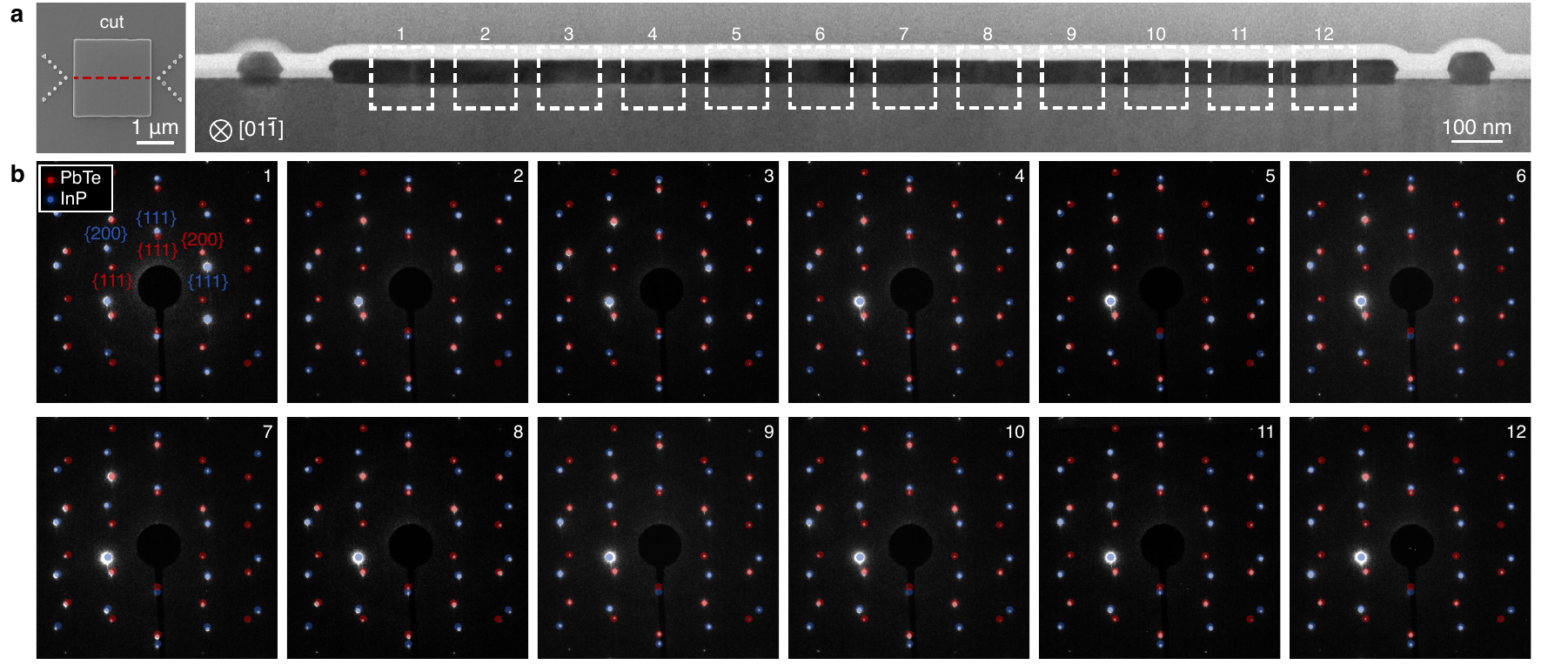}
	\caption{\textbf{Single crystalline PbTe SAG. 
	a,}~Top-view SEM image of selective area grown PbTe in a $2\times2~\upmu$m opening on InP~(111)A. The location from which the cross-sectional TEM lamella is taken is indicated by a red dashed line. The right side shows a bright-field TEM micrograph of the cross-section. Unlike the image shown in Fig.~4a, it is taken in zone-axis of the PbTe crystal.
    \textbf{b,}~Electron diffraction patterns of the twelve regions marked in panel~\textbf{a}. The entire PbTe structure is single crystalline with a twinned epitaxial relation between InP and PbTe.}
\end{figure*}

\clearpage
\begin{figure}
	\centering
	\includegraphics[width=\textwidth]{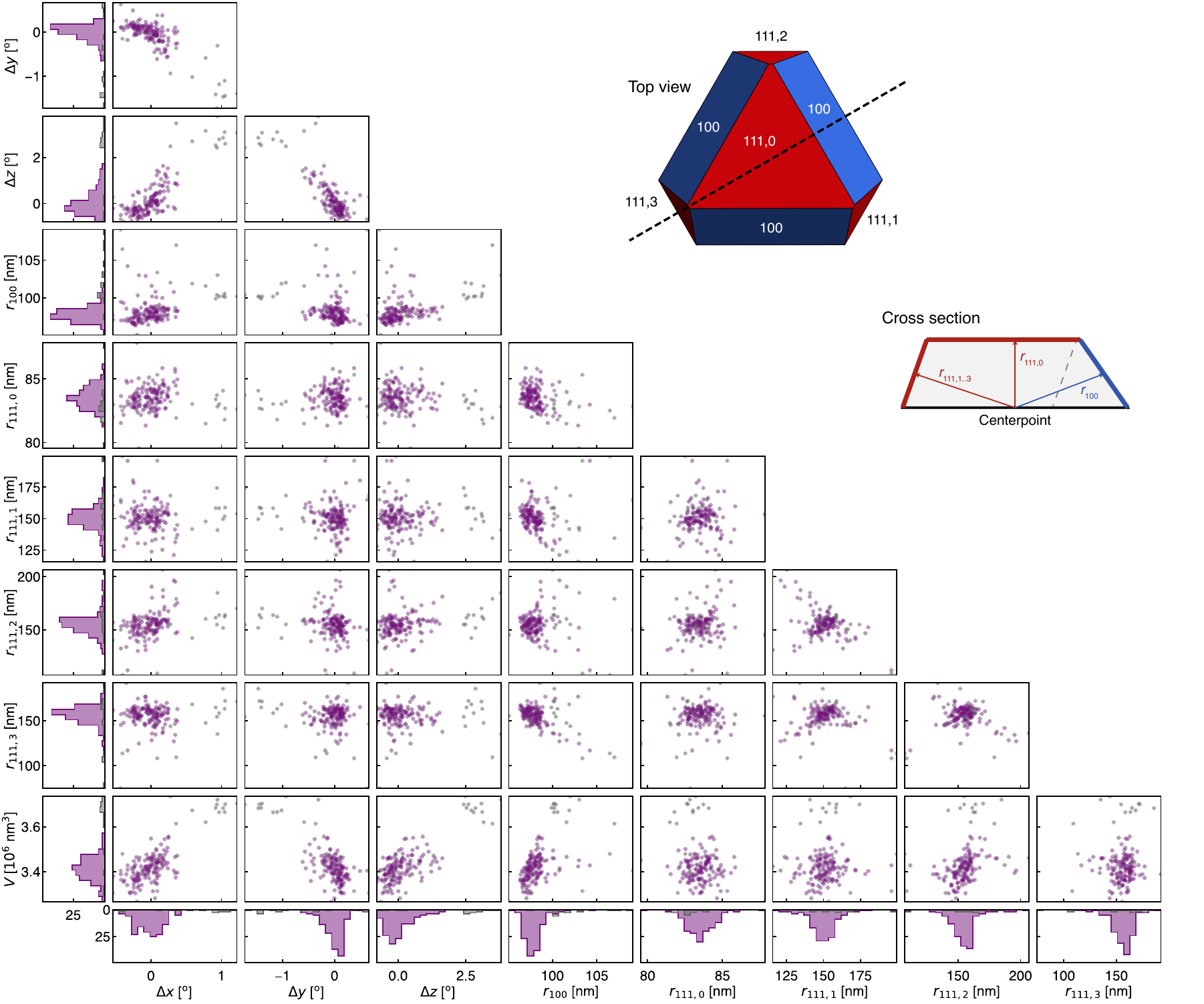}
	\caption{\textbf{Crystal fitting data.} Correlation matrix for all fitting parameters used to describe the PbTe crystals shown in Fig.~4c and d. In addition, the crystal volume $V$ is included. Bad fits with a relative residual volume above 4\% are plotted in grey. The leftmost column and the bottom row contain the histograms of the quantity shown on that axis in that row and column, respectively. Schematics in the top right corner indicate the fitting parameter $r_{100}$ and $r_{111,0-3}$. The plane of the cross-section is indicated by a black dashed line. The angular data shows faint signatures of cross-correlation. The facet radii show no correlation between either families or to the structure misorientation. This indicates that misorientation neither promotes nor hampers the growth of individual facets or the structures as a whole.} 
\end{figure}

\clearpage
\bibliographystyle{naturemag}
\bibliography{ref.bib}